\documentstyle[epsfig]{aprim}

\newif\ifAMStwofonts

\ifoldfss
  \ifCUPmtlplainloaded \else
    \NewTextAlphabet{textbfit} {cmbxti10} {}
    \NewTextAlphabet{textbfss} {cmssbx10} {}
    \NewMathAlphabet{mathbfit} {cmbxti10} {} 
    \NewMathAlphabet{mathbfss} {cmssbx10} {} 
  \fi
  \ifAMStwofonts
    \ifCUPmtlplainloaded \else
      \NewSymbolFont{upmath} {eurm10}
      \NewSymbolFont{AMSa} {msam10}
      \NewMathSymbol{\upi}     {0}{upmath}{19}
      \NewMathSymbol{\umu}     {0}{upmath}{16}
      \NewMathSymbol{\upartial}{0}{upmath}{40}
      \NewMathSymbol{\leqslant}{3}{AMSa}{36}
      \NewMathSymbol{\geqslant}{3}{AMSa}{3E}

    \fi
  \fi
\fi 

\ifnfssone
  \newmathalphabet{\mathit}
  \addtoversion{normal}{\mathit}{cmr}{m}{it}
  \addtoversion{bold}{\mathit}{cmr}{bx}{it}
  \newmathalphabet{\mathbfit} 
  \addtoversion{normal}{\mathbfit}{cmr}{bx}{it}
  \addtoversion{bold}{\mathbfit}{cmr}{bx}{it}
  \newmathalphabet{\mathbfss} 
  \addtoversion{normal}{\mathbfss}{cmss}{bx}{n}
  \addtoversion{bold}{\mathbfss}{cmss}{bx}{n}
  \ifAMStwofonts
    \ifCUPmtlplainloaded \else
      \UseAMStwoboldmath
      \makeatletter
      \new@mathgroup\upmath@group
      \define@mathgroup\mv@normal\upmath@group{eur}{m}{n}
      \define@mathgroup\mv@bold\upmath@group{eur}{b}{n}
      \edef\UPM{\hexnumber\upmath@group}
      \new@mathgroup\amsa@group
      \define@mathgroup\mv@normal\amsa@group{msa}{m}{n}
      \define@mathgroup\mv@bold\amsa@group{msa}{m}{n}
      \edef\AMSa{\hexnumber\amsa@group}
      \makeatother
      \mathchardef\upi="0\UPM19
      \mathchardef\umu="0\UPM16
      \mathchardef\upartial="0\UPM40
      \mathchardef\leqslant="3\AMSa36
      \mathchardef\geqslant="3\AMSa3E
    \fi
  \fi
\fi 

\ifnfsstwo
  \DeclareMathAlphabet{\mathbfit}{OT1}{cmr}{bx}{it}
  \SetMathAlphabet\mathbfit{bold}{OT1}{cmr}{bx}{it}
  \DeclareMathAlphabet{\mathbfss}{OT1}{cmss}{bx}{n}
  \SetMathAlphabet\mathbfss{bold}{OT1}{cmss}{bx}{n}
  \ifAMStwofonts
    \ifCUPmtlplainloaded \else
      \DeclareSymbolFont{UPM}{U}{eur}{m}{n}
      \SetSymbolFont{UPM}{bold}{U}{eur}{b}{n}
      \DeclareSymbolFont{AMSa}{U}{msa}{m}{n}
      \DeclareMathSymbol{\upi}{0}{UPM}{"19}
      \DeclareMathSymbol{\umu}{0}{UPM}{"16}
      \DeclareMathSymbol{\upartial}{0}{UPM}{"40}
      \DeclareMathSymbol{\leqslant}{3}{AMSa}{"36}
      \DeclareMathSymbol{\geqslant}{3}{AMSa}{"3E}
    \fi
  \fi
\fi 

\ifCUPmtlplainloaded \else
  \ifAMStwofonts \else 
    \def\upi{\pi}
    \def\umu{\mu}
    \def\upartial{\partial}
  \fi
\fi

\title[Manuscript Template]{The Evolving Structure of Galactic Disks}

\author[Martel et al.]
       {Hugo Martel$^1$, Chris Brook$^1$, Sean McGee$^{1,2}$,
        Brad Gibson$^3$, \& Daisuke Kawata$^{3,4}$\\
        $^1$D\'epartment de physique, g\'enie physique et optique, 
            Universit\'e Laval, Qu\'ebec, Canada\\
        $^2$University of Waterloo, Waterloo, Canada\\
        $^3$Swinburne University, Melbourne, Australia\\
        $^4$Carnegie Observatories, Washington, United States}
\date{}

\pagerange{\pageref{firstpage}--\pageref{lastpage}}
\pubyear{2005}

\begin{document}

\maketitle

\label{firstpage}

\begin{abstract}
Observations suggest that the structural parameters of disk galaxies have
not changed greatly since redshift 1. We examine whether these observations
are consistent with a cosmology in which structures form hierarchically.
We use SPH/N-body galaxy-scale simulations
to simulate the formation and evolution of Milky-Way-like
disk galaxies by fragmentation, followed by hierarchical merging. The
simulated galaxies have a thick disk, that forms in a period of chaotic
merging at high redshift, during which a large amount of $\alpha$-elements
are produced, and a thin disk, that forms later and has a higher metallicity.
Our simulated disks settle down quickly and do not evolve much since
redshift $z\sim1$, mostly because no major mergers take place between
$z=1$ and $z=0$. During this period, the disk radius increases (inside-out
growth) while its thickness remains constant. These results are consistent
with observations of disk galaxies at low and high redshift.
\end{abstract}

\begin{keywords}
  cosmology --- galaxies: formation --- galaxies: evolution
\end{keywords}

\section{The Simulations}

We used the chemodynamical galaxy formation code GCD+ \cite{kg03}, to simulate
the formation of 4 disk galaxies of masses comparable to the mass of the 
Milky Way. The code includes self-gravity, hydrodynamics, radiative cooling,
star formation, supernova feedback, and metal enrichment. The initial 
conditions consist of a uniform, slowly-rotating sphere of gas of mass
$5\times10^{11}M_\odot$ onto which we superpose small density fluctuations.
The systems initially fragment into several clumps that later collide
and merge to form disk galaxies with 3 distinct structures: a thin disk,
a thick disk, and a halo.

\section{Disk Formation and Growth}

Figure 1 shows the age of stars vs. their height above the galactic plane
and their distances from the galactic center, for one simulated galaxy.
This reveals the existence of two distinct disks. The thin disk is about
2~kpc thick, and is composed of old stars ($>9$~Gyrs). Both disks extend
to a radius of 10~kpc \cite{brook05a}.

\begin{figure}
 \begin{center}
   \epsfig{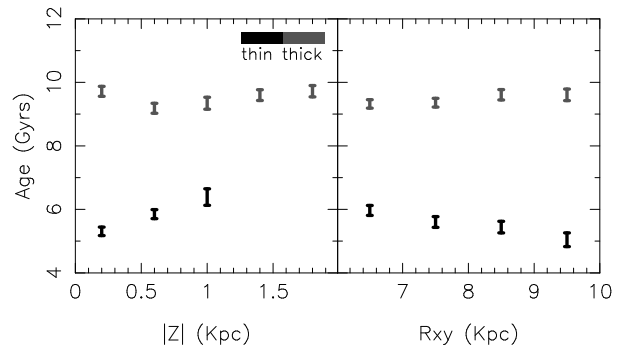}
 \end{center}
 \caption{Age vs. height and galactic radius, for the thin disk (lower symbols)
          and the thick disk (upper symbols). \label{fig1}}
\end{figure}

Figure 2 shows the scale height $h_z$ and scale length $h_l$ of our
4 simulated galaxies, at 3 different epochs. The time sequence (stars
$\rightarrow$ squares $\rightarrow$ triangle) shows that $h_l$
tends to increase with time while $h_z$ remains constant, indicating that
disks grow ``inside-out'' \cite{brook05b}. The time sequence
(circles $\rightarrow$ crosses), based on observations at
$z=1$ and $z=0$ \cite{sd00,rdc03} is consistent with that result.

\begin{figure}
 \begin{center}
   \epsfig{file=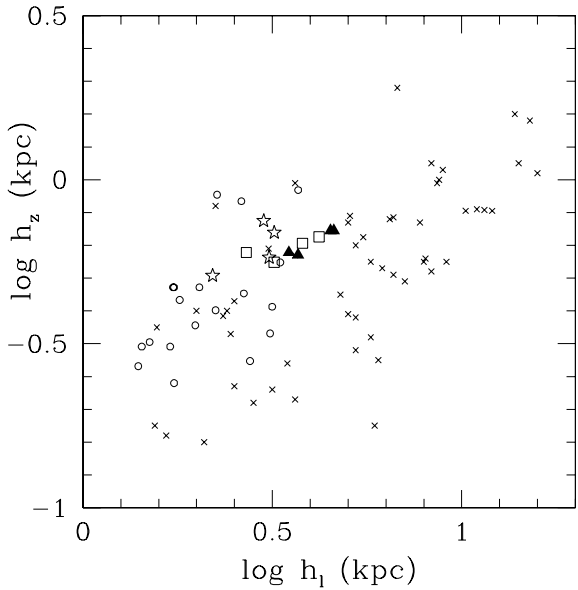,width=6.0cm}
 \end{center}
 \caption{Scale-height $h_z$ vs. scale length $h_l$ for our 4 simulated
          galaxies, shown at redshifts $z=1$ (stars), $z=0.5$ (squares),
          and $z=0$ (triangles). Also shown are observations of local
          disk galaxies (crosses) and edge-on galaxies at
          $z\sim1$ (circles)}.
\end{figure}

\section{Abundance Patters}

Figure 3 shows the abundances of $\alpha$-elements (O, Mg, and Si) vs.
metallicity. The stars in the thick disk and halo form early. They are
rich in $\alpha$-elements but poor in iron compared to the thin disk.
The high abundance of $\alpha$-elements results from the high occurrence
of Type II supernovae at early time, during the epoch of chaotic merging.

\begin{figure}
 \begin{center}
   \vskip-0.1in
   \epsfig{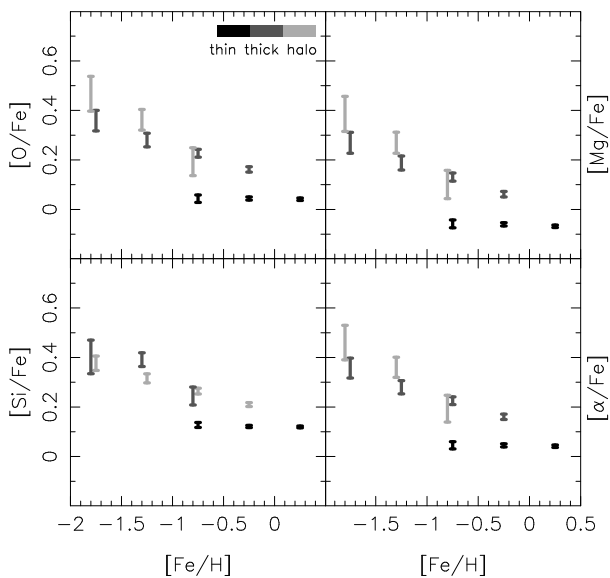}
 \end{center}
 \caption{Abundance of $\alpha$-elements vs. metallicity, for stars in the
thin disk (lower symbols), thick disk (middle symbols), 
and halo (upper symbols). The last panel
shows the combined abundance of oxygen, magnesium, and silicon.
\label{fig3}}
\end{figure}

Figure 4 shows the gradients of metallicity. The metallicity of the thick disk
tends to decrease with both height and distance from the center. For the thin
disk, the metallicity decreases with height but does not vary with distance.
There is a clear separation between the two components, the thick
disk being significantly metal-poor compared to the thin disk (even though
it is richer in $\alpha$-elements). Figure 5 shows a histogram of the
number of stars vs. metallicity. The majority of stars are in the thin disk,
and these stars have a higher metallicity than the thick disk or halo stars.

\begin{figure}
 \begin{center}
   \epsfig{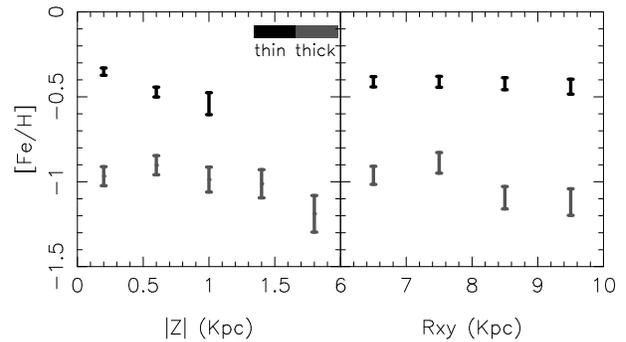}
 \end{center}
 \caption{Metallicity vs. height and galactic radius, for the thin disk
          (upper symbols) and thick disk (lower symbols).
\label{fig4}}
\end{figure}

\begin{figure}
 \begin{center}
   \epsfig{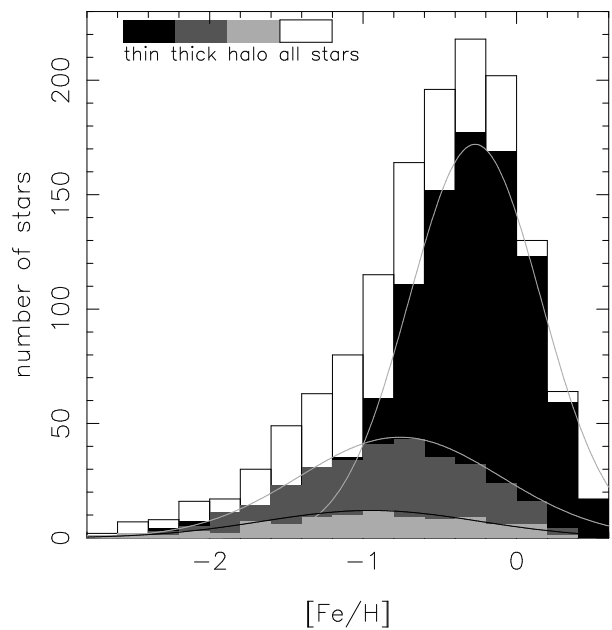}
 \end{center}
 \caption{Number of stars vs.metallicity, for the thin disk (black),
thick disk (dark gray), halo (light gray), and total (white).
Each ``star'' is actually
a computational object that represents $\sim26,000$ stars.
\label{fig5}}
\end{figure}

\bigskip

\noindent
This work was supported by the Natural Science and Engineering 
Research Council of Canada. We are very thankful to Vincent
Veilleux for producing several of the figures.

\clearpage

\end{document}